\documentclass[letterpaper,twocolumn,prl,aps,superscriptaddress,amsmath,amssymb,floatfix]{revtex4-2}
\usepackage{mathptmx}
\usepackage[latin9]{inputenc}
\usepackage{color}
\usepackage{upgreek}
\usepackage{amsmath}
\usepackage{amssymb}
\usepackage{calrsfs}
\DeclareMathAlphabet{\pazocal}{OMS}{zplm}{m}{n}
\SetMathAlphabet\pazocal{bold}{OMS}{zplm}{bx}{n}
\usepackage{graphicx}
\usepackage{esint}
\usepackage[hypertexnames=false, breaklinks=true, bookmarksnumbered=true, bookmarksopen=true, colorlinks=true, linktocpage=true, citecolor=blue, urlcolor=magenta, linkcolor=magenta]{hyperref}
\makeatletter

\pdfpageheight\paperheight
\pdfpagewidth\paperwidth

\usepackage{textcomp}
\usepackage{epstopdf}

\pdfpageheight\paperheight
\pdfpagewidth\paperwidth

\@ifundefined{textcolor}{}{%
 \definecolor{BLACK}{gray}{0}
 \definecolor{WHITE}{gray}{1}
 \definecolor{RED}{rgb}{1,0,0}
 \definecolor{GREEN}{rgb}{0,1,0}
 \definecolor{BLUE}{rgb}{0,0,1}
 \definecolor{CYAN}{cmyk}{1,0,0,0}
 \definecolor{MAGENTA}{cmyk}{0,1,0,0}
 \definecolor{YELLOW}{cmyk}{0,0,1,0}
}

\usepackage{xcolor}\usepackage{soul}
\makeatother

\begin{document}
\title{Interaction-Enhanced Superradiance of a Ryderg-Atom Array}
\author{Yiwen Han}
\affiliation{CAS Key Laboratory of Quantum Information, University of Science and Technology of China, Hefei 230026, China}
\author{Haowei Li}
\affiliation{CAS Key Laboratory of Quantum Information, University of Science and Technology of China, Hefei 230026, China}
\author{ Wei Yi}
\email{wyiz@ustc.edu.cn}
\affiliation{CAS Key Laboratory of Quantum Information, University of Science and Technology of China, Hefei 230026, China}
\affiliation{CAS Center For Excellence in Quantum Information and Quantum Physics, Hefei 230026, China}
\affiliation{Hefei National Laboratory, University of Science and Technology of China, Hefei 230088, China}
\date{\today}
\begin{abstract}
We study the superradiant phase transition of an array of Rydberg atoms in a dissipative microwave cavity. Under the interplay of the cavity field and the long-range Rydberg interaction, the steady state of the system exhibits an interaction-enhanced superradiance,
with vanishing critical atom-cavity coupling rates at a discrete set of interaction strengths. We find that, while the phenomenon can be analytically understood in the case of a constant all-to-all interaction, the enhanced superradiance persists under typical experimental parameters with spatially dependent interactions, but at modified critical interaction strengths. The diverging susceptibility at these critical points is captured by emergent quantum Rabi models, each of which comprises a pair of collective atomic states with different numbers of atomic excitations. These collective states become degenerate at the critical interaction strengths, resulting in a superradiant phase for an arbitrarily small atom-cavity coupling.
\end{abstract}
\maketitle

\emph{Introduction.}
Combining strong long-range interaction, flexible spatial configuration, and large polarizability, Rydberg-atom arrays are emerging as an increasingly important platform for quantum simulation and computation~\cite{rdbRev2,rdbRev3,rdbRev4,rdbRev1}. During the past decade, a vast range of intriguing many-body phenomena have been simulated and investigated in Rydberg-atom arrays, including magnetism and dynamics in quantum spin models~\cite{Leseleuc18,Bernien17,Bijnen15,Zeiher17,Scholl21}, symmetry protected topological phases~\cite{Leseleuc19}, coherent excitation transfer~\cite{Barredo15,Lienhard20,Han24}, as well as emergent gauge field~\cite{gauge1,gauge2,gauge3,gauge4} and many-body localization~\cite{Machado21,Marcuzzi17}.
Furthermore, the recent demonstration of error suppression using logical qubits on reconfigurable Rydberg-atom arrays represents a seminal first step toward an era of error-corrected intermediate-size quantum systems~\cite{Bluvstein24}.
On the other hand, while the Rydberg states are relatively long-lived, dissipation in Rydberg atoms is further tunable through a variety of means, including laser-induced loss, microwave dressing, or coupling to a cavity. As such, Rydberg atoms are also ideal for the study of dissipative many-body dynamics~\cite{cross12,Carr13,Lesanovsky13,Mal14,ott17,helnat,Li24}.
A particularly interesting configuration is the atom-cavity hybrid system involving Rydberg states~\cite{Zhang13,Yan15}, where the cavity-induced long-range interatomic interactions and the back action of cavity dissipation conspire with the Rydberg interactions, leading to exotic non-equilibrium dynamics.

For atoms coupled to a cavity, an outstanding phenomenon is the superradiant phase transition, which can be traced back to the study of the Dicke model half a century ago~\cite{Dicke54,Hepp73}.
In the Dicke model, an ensemble of two-level, non-interacting atoms are coupled to a single-mode electromagnetic field, corresponding to the atom-cavity coupling in the aforementioned hybrid system.
A superradiant phase transition occurs beyond a critical atom-cavity coupling strength, wherein the cavity field becomes macroscopically populated~\cite{Hepp73,Wang73,ssRev1,ssRev2}.
While the superradiant transition also arises in open quantum systems~\cite{Torre13,Padi23,Chen06}, the advent of ultracold quantum gases further enriches its study~\cite{Nagy10,Nagy11,Chen14,keeling14,piazza14}.
In particular, the transition was observed in a Bose-Einstein condensate of atoms coupled to a dissipative cavity field, where, driven by the collective light-matter interaction, the macroscopic population of the steady-state cavity field is accompanied by a self-organized density pattern in the condensate~\cite{Baumann10}.
By contrast, a Fermi-surface-nesting-enhanced superradiance was predicted and subsequently observed for degenerate Fermi gases in a cavity~\cite{fermisr}.
In these studies, the interatomic interactions are weak at best. Whether atomic interactions, particularly the long-range interactions of Rydberg states, can have a significant impact on the steady-state superradiant transition, remains to be explored.

\begin{figure}[tbp]
\begin{centering}
\includegraphics[width=\linewidth]{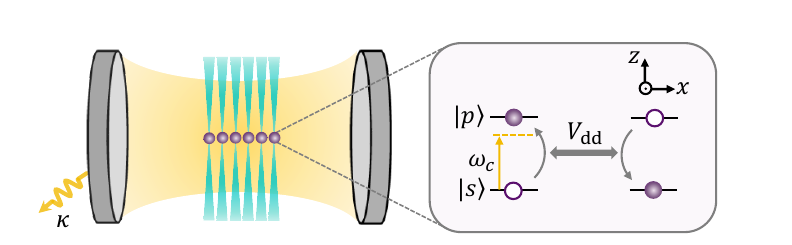}
\par\end{centering}
\caption{Schematic illustration of Rydberg atoms in a microwave cavity. The green stripes indicate the optical tweezers, used for trapping and arranging the atom array. The dissipative cavity, characterized by a decay rate $\kappa$, off-resonantly couples two Rydberg sates $|s\rangle\leftrightarrow |p\rangle$. Microscopic illustration of the resonant dipole-dipole interaction $V_{\text{dd}}$ is shown to the right.}
\label{Fig1}
\end{figure}

In this work, we reveal a Rydberg-interaction-enhanced superradiance in an array of atoms with microwave-cavity assisted Rydberg-state coupling. By studying the steady-state superradiance of the atom-cavity hybrid system, we show that the critical atom-cavity coupling rate for the superradiant transition vanishes at a series of discrete interaction strengths, suggesting divergent susceptibility at these critical points. Invoking a simplified model with a constant all-to-all interaction, we derive an analytic expression for the critical points, and find that the divergence of susceptibility thereof coincides with the two-fold degeneracies of symmetric collective states in the subspaces with $n$ and $n + 1$ excitations, respectively. We then demonstrate, through numerical simulations, that the interaction-enhanced superradiance persists under typical experimental conditions, where both the Van der Waals interaction and the spatial dependence of these interactions are considered. Compared to the constant-interaction case, the critical interaction strengths are modified, but a similar set of degenerate collective states also play a key role. Notably, these collective states are no longer symmetric due to the spatial dependence of the interaction potential. In either case, we show that the enhanced superradiance near a critical point can be captured by an emergent quantum Rabi model, which comprises the relevant near-degenerate collective states.

\emph{System configuration.}
We consider a one dimensional Rydberg-atom array with a lattice constant $R_0$, confined by optical tweezers in a single-mode microwave cavity along the horizontal $x$ direction, as illustrated in Fig.~\ref{Fig1}. For each atom, we consider two Rydberg states, denoted respectively as $\left|s\right\rangle$ and $\left|p\right\rangle$. The Rydberg states are coupled by the cavity mode characterized by frequency $\omega_{c}$ and wavelength $\lambda_{c}$. In the case that $R_0$ is much smaller than $\lambda_{c}$, the cavity field is identical for all atoms, leading to a homogeneous atom-cavity coupling rate $\pazocal{G}$.
The Rydberg atoms interact through the resonant dipole interaction
$V\left(R_{jk}\right)=C_3/R_{jk}^{3}$, where $C_3$ is the dipole-dipole interaction parameter, and $R_{jk}$ is the distance between the $j$ th and $k$ th atoms along the chain.
Note that, although the Van der Waals interactions become stronger than the dipole interaction at short interatomic distances, they do not qualitatively modify the interaction-enhanced superradiance (see later discussions). We therefore neglect the Van der Waals interactions for now.

We start by studying the simplified case where the dipole-dipole interaction is approximated by a
constant all-to-all interaction $V\left(R_{jk}\right)=V_{\text{dd}}$.
The simplified atom-cavity coupling Hamiltonian is a generalized Dicke model with dipole interactions (we take $\hbar=1$)~\cite{Lawande85,Chen06,Dou22,Zhang23}
\begin{equation}
\hat{H}=\hat{H}_{\text{atom}}+\omega_{c}\hat{a}^{\dagger}\hat{a}+\sqrt{\frac{2}{N}}\pazocal{G}\sum_{j=1}^{N}\left(\hat{a}^{\dagger}+\hat{a}\right)\hat{\sigma}_{j}^{x},
\label{eq:H}
\end{equation}
where the atomic Hamiltonian
\begin{equation}
\hat{H}_{\text{atom}}=\frac{\omega_{a}}{2}\sum_{j=1}^{N}\hat{\sigma}_{j}^{z}+\sum_{j<k}V_{\text{dd}}\left(\hat{\sigma}_{j}^{-}\hat{\sigma}_{k}^{+}+\text{H.c.}\right).
\end{equation}
Here $N$ and $\omega_{a}$ are the total atom number and the energy-level difference between the two Rydberg states, respectively, $\hat{a}$ and $\hat{a}^\dag$ are the annihilation and creation operators of the cavity photon, and $\hat{\sigma}_{j}^{\pm}=\hat{\sigma}_{j}^{x}\pm i\hat{\sigma}_{j}^{y}$, and $\hat{\sigma}_{j}^{\beta}$ ($\beta=x,y,z$) are the Pauli operators for the $j$ th atom, defined through $\hat{\sigma}^+_j=|p\rangle\langle s|_j$, $\hat{\sigma}^-_j=|s\rangle\langle p|_j$, $\hat{\sigma}^z_j=|p\rangle\langle p|_j-|s\rangle\langle s|_j$, with $|p\rangle_j$ and $|s\rangle_j$ indicating the Rydberg states of the $j$ th atom.

Importantly, we consider an off-resonant coupling between the microwave cavity field and the inter-Rydberg-state transition $|s\rangle\leftrightarrow |p\rangle$,
so that the counter-rotating terms are retained in Eq.~(\ref{eq:H}).
Further, since the dependence of the Hamiltonian on the atomic degrees of freedom is fully accounted for by the collective spin operators $\hat{S}^{\beta}=\sum_{j}\hat{\sigma}_{j}^{\beta}/2$ and $\hat{S}^{\pm}=\sum_{j}\hat{\sigma}_{j}^{\pm}$, we can rewrite Hamiltonian (\ref{eq:H}) as
\begin{equation}
\hat{H}=\omega_{a}\hat{S}^{z}+V_{\text{dd}}\hat{S}^{+}\hat{S}^{-}+\omega_{c}\hat{a}^{\dagger}\hat{a}+2\sqrt{\frac{2}{N}}\pazocal{G}\left(\hat{a}^{\dagger}+\hat{a}\right)\hat{S}^{x}.
\label{eq:HS}
\end{equation}

In a realistic system, dissipation is inevitable through its coupling to the environment. For our setup, the lifetime of the Rydberg states are enhanced as the cavity mode suppresses the effects of the blackbody radiation on the Rydberg states. We further assume a negligible spontaneous decay rate from the Rydberg states $|p\rangle$ and $|s\rangle$ to the electronic ground state, so that the dominant dissipation channel is the cavity decay. The dynamics of the atom-cavity hybrid system is then governed by the Lindblad master equation
\begin{equation}
\dot{\hat{\rho}}=-i\left[\hat{H},\hat{\rho}\right]+\kappa\left(2\hat{a}\hat{\rho}\hat{a}^{\dagger}-\hat{a}^{\dagger}\hat{a}\hat{\rho}-\hat{\rho}\hat{a}^{\dagger}\hat{a}\right),
\label{eq:lindblad}
\end{equation}
where $\hat{\rho}$ is the density matrix containing both the atomic and cavity-photon degrees of freedom, and $\kappa$ is the cavity decay rate.

\emph{Interaction-enhanced superradiance.}
We focus on the superradiant transition in the steady state of the atom-cavity hybrid system.
Adopting a mean-field approximation on the cavity field,
we replace $\hat{a}$ with the expectation value $\alpha=\langle\hat{a}\rangle$, which, under the stationary condition $\partial \alpha/\partial t= 0$, is given by
\begin{equation}
\alpha=\frac{-i2\sqrt{\frac{2}{N}}\pazocal{G}}{i\omega_{c}+\kappa}\left\langle \hat{S}^{x}\right\rangle.\label{eq:alpha}
\end{equation}
The steady state is solved self-consistently by numerically diagonalizing Hamiltonian (\ref{eq:HS}) while enforcing Eq.~(\ref{eq:alpha}).

\begin{figure}[tbp]
\begin{centering}
\includegraphics[width=1\linewidth]{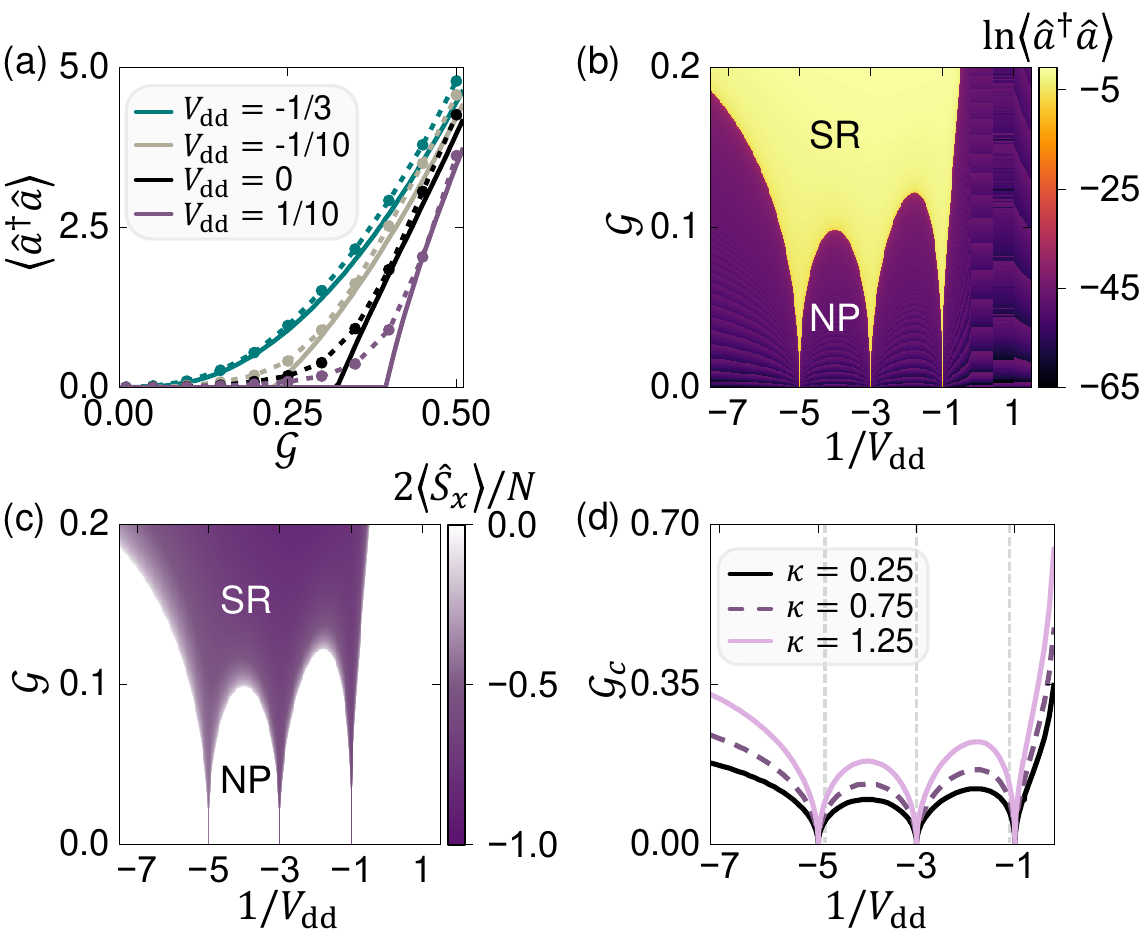}
\par\end{centering}
\caption{(a) The steady-state cavity photon number as a function of the atom-cavity coupling rate for different interaction strengths. Solid lines represent results under the mean-field approximation, while dash-dotted lines are obtained by evolving the quantum master equation (\ref{eq:lindblad}) for sufficiently long times, starting from the initial direct-product state $\left | ss\dots s\right\rangle\otimes\left |0\right\rangle$ of the atom-cavity hybrid system. Here $|0\rangle$ is the vacuum state of cavity photons. Different colors correspond to different interaction strengths. (b)(c) Steady-state phase diagrams, characterized by the cavity photon number and the collective atom correlation $\hat{S}_{x}$, respectively. Here ``SR'' stands for the superradiant phase, and ``NP'' stands for the normal phase.
(d) Impact of the cavity dissipation on the critical atom-cavity coupling rate of the superradiant transition. Different colors and line shapes correspond to different cavity decay rates. For all calculations, we take $\omega_a$ as the unit of energy, with other parameters given by $N=6$, $\omega_c=0.75$, and $\kappa=0.25$ [for (a)(b)(c)].}
\label{Fig2}
\end{figure}

In Fig.~\ref{Fig2}(a), we show the numerically calculated steady-state cavity photon number $\langle\hat{a}^\dag \hat{a}\rangle=|\alpha|^2$ (solid curves) with increasing atom-cavity coupling strength $\pazocal{G}$, for an array of $N=6$ atoms.
Under most of the interaction strengths, as well as in the non-interacting case, a superradiant transition can be identified at a critical $\pazocal{G}_c$, across which the cavity field starts to become finite. Importantly, the sign and strength of the Rydberg interaction $V_{\text{dd}}$ have a significant impact on the superradiant transition. As a dramatic example, at $V_{\text{dd}}=-1/3$, the critical $\pazocal{G}_c$ vanishes---the system becomes superradiant even for an arbitrarily small atom-cavity coupling rate. The mean-field results above are consistent with those from a full-quantum calculation [see dashed curves in Fig.~\ref{Fig2}(a)], for which we evolve the Lindblad equation (\ref{eq:lindblad}) for a sufficiently long time so that the system is close to the steady state.
While the superradiant phase transitions from the full-quantum treatment are not as sharp compared to the mean-field results, at the critical value of  $V_{\text{dd}}=-1/3$, both calculations indicate a vanishing $\pazocal{G}_c$.

The overall interaction dependence of the superradiant transition is more transparent in Fig.~\ref{Fig2}(b)(c), where the color contour respectively indicates the steady-state cavity photon number and the collective atom correlation $\langle\hat{S}_x\rangle$, on the $1/V_{\text{dd}}$--$\pazocal{G}$ plane.
Consistent with the collective nature of superradiance, the macroscopic population of the cavity field is always accompanied by the emergence of collective atomic correlations.
Furthermore, we find that repulsive interactions ($V_{\text{dd}}>0$) tend to suppress the transition, leading to an ever increasing $\pazocal{G}_c$ for larger $V_{\text{dd}}$. By contrast, under attractive interactions ($V_{\text{dd}}<0$), $\pazocal{G}_c$ exhibits an undulating pattern. It vanishes at some discrete integer values of $1/V^c_{\text{dd}}$, indicating interaction-enhanced superradiance, as first observed in Fig.~\ref{Fig2}(a).
Furthermore, while the critical coupling strength $\pazocal{G}_c$ generally depends on the cavity-decay rate $\kappa$, the critical interaction strengths $V^c_{\text{dd}}$ where $\pazocal{G}_c$ vanishes are independent of the cavity decay [see Fig.~\ref{Fig2}(d)].
While such independence has potential significance for engineering superradiant transitions in bad cavities, it also strongly suggests an underlying mechanism hinged upon the atomic correlations.

\emph{An emergent quantum Rabi model.}
In the standard Dicke model, the superradiant transition can be analyzed by dividing the Hilbert space into subspaces with different numbers of atomic excitations, and the transition is found to be dominated by symmetric states in each subspace~\cite{Bull87,supp}. Of particular importance are the ground state and the first excited state, the former with no atomic excitations and the latter a symmetric superposition of all instances of a single excitation.
The critical atom-cavity coupling rate of the superradiant transition is related to the energy difference between these two states.
Now the addition of Rydberg interactions would modify the spectral outlook of the collective atomic excitations, giving rise to not only level shifts but also degeneracies in the low-lying collective states.
As we show below, it is this interaction-induced low-energy degeneracy that is directly responsible for the interaction-enhanced superradiance.

\begin{figure}[tbp]
\begin{centering}
\includegraphics[width=1\linewidth]{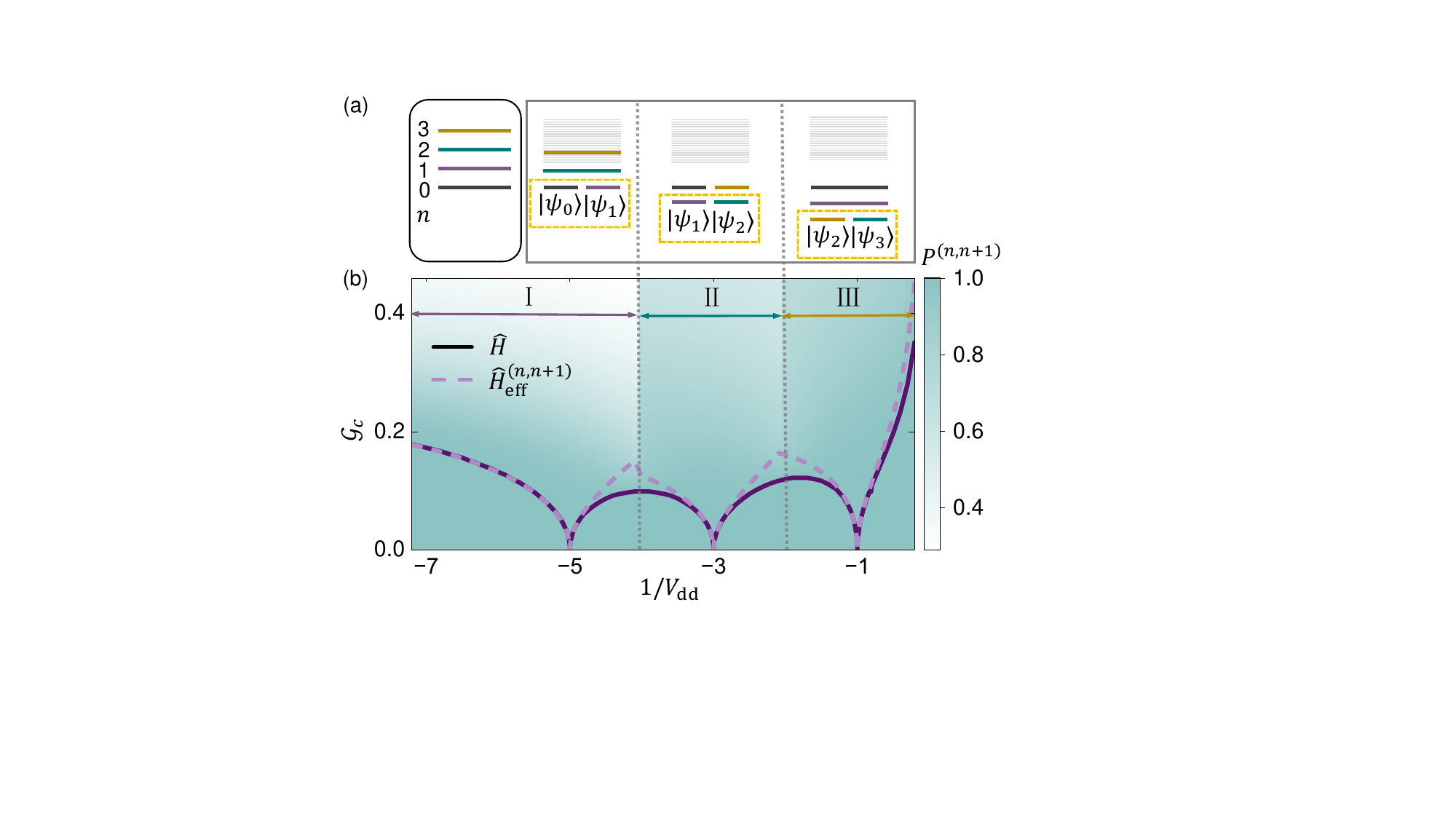}
\par\end{centering}
\caption{(a) Schematic illustration of the energy shift and degeneracies of different symmetric states (bold lines) with different interaction strengths. The left subplot illustrates the energy levels of the Dicke model with $V_{\text{dd}} =0$. From bottom to top, the colored bold lines correspond to subspaces with $n=0,1,2,3$, respectively, where collective states in the same subspace have the same energy. The subplot on the right illustrates the energy levels at the critical interaction strengths $1/V^c_{\text{dd}}=-5,-3,-1$ (left, middle, right), respectively. Here the bold lines indicate the energy levels of the symmetric states, their colors indicate different excitation numbers, consistent with those in the case of $V_{\text{dd}} =0$ (left subplot).
The thin grey lines denote non-symmetric collective states.
(b) The critical atom-cavity coupling rate for the superradiant transition with varying interaction strengths, calculated from the effective quantum Rabi models (dashed lines) and the full Hamiltonian (solid lines), respectively (see main text for detailed discussion). The three regions (marked with I, II, and III), separated by the vertical dotted lines at $1/V_{\text{dd}}=-4,-2$, are centered at the critical interaction strengths
$1/V^c_{\text{dd}}=-5,-3,-1$.
The color contour in the background represents the overlap $P^{\left(n,n+1\right)}$ between the steady state and the degenerate ground-state manifold at the central critical interaction strength $V^c_{\text{dd}}$ of the corresponding region. Other parameters and the unit of energy are the same as those in Fig.~\ref{Fig2}(b)(c).}
\label{Fig3}
\end{figure}

With the physical understanding above, we define the states $\{|\psi_n\rangle\}$, which are the symmetric superpositions of all possible $n$-atom excitations, with
\begin{equation}
|\psi_n\rangle=\frac{1}{\sqrt{C_{N}^{n}}}\sum_{\{j\dots k\}}|s\dots p_{j}\dots p_{k} \dots s\rangle,
\label{eq:psi}
\end{equation}
where $\{j\dots k\}$ in the summation denotes all possible combinations of $n$ excitation locations.
In the absence of the cavity field, the corresponding eigenenergies for these collective states are $\omega_{n}=-\left(\frac{N}{2}-n\right)\omega_{a}+n\left(N-n\right)V_{\text{dd}}$. As illustrated in Fig.~\ref{Fig3}(a), with increasing attractive interaction $V_{\text{dd}}$, the energies of some symmetric excited states shift downward significantly, and the ground state becomes degenerate with at a discrete set of interaction strengths
\begin{equation}
V^c_{\text{dd}}(n)=-\frac{\omega_{a}}{N-\left(2n+1\right)},
\label{eq:Vdd}
\end{equation}
where $n\leq N/2$. More specifically, at each of the critical interaction strength $V^c_{\text{dd}}(n)$, the two-fold degenerate ground-state subspace is spanned by $|\psi_{n}\rangle$ and $|\psi_{n+1}\rangle$. It follows that the superradiant transition near $V^c_{\text{dd}}(n)$ is dominated by the two near-degenerate states $|\psi_{n}\rangle$ and $|\psi_{n+1}\rangle$.
To quantitatively confirm this point, we divide the overall range of $1/V_{\text{dd}}$ into different regions, centered at the critical points with vanishing $\pazocal{G}_{c}$. These regions are separated by the vertical dotted lines in Fig.~\ref{Fig3}(b).
As the background of Fig.~\ref{Fig3}(b), we show the color contour of the overlap $P^{\left(n,n+1\right)}$ between the steady state (denoted as $|\psi_s\rangle$) and the ground-state manifold at the critical $V^c_{\text{dd}}$ of the corresponding region. For instance, we consider $\{|\psi_0\rangle,|\psi_1\rangle\}$ in region I, $\{|\psi_1\rangle,|\psi_2\rangle\}$ in II, $\{|\psi_2\rangle,|\psi_3\rangle\}$ in III, and
$P^{\left(n,n+1\right)}=\left|\langle\psi_{n}|\psi_{s}\rangle\right|^2+\left|\langle\psi_{n+1}|\psi_{s}\rangle\right|^2$
in the corresponding region.
Apparently, $P^{\left(n,n+1\right)}$ is close to unity in the broad vicinity of the superradiant phase transition.

The observation above inspires us to examine an emergent quantum Rabi model
\begin{align}
 \hat{H}_{\text{eff}}^{\left(n,n+1\right)}=&\frac{\Delta_{n}}{2}\left(\left|\psi_{n+1}\right\rangle \left\langle \psi_{n+1}\right|-\left|\psi_{n}\right\rangle \left\langle \psi_{n}\right|\right)+\omega_{c}\hat{a}^{\dagger}\hat{a}\notag\\
 &+\sqrt{\frac{2}{N}}\pazocal{G}\eta_{n}\left(\hat{a}^{\dagger}+\hat{a}\right)\left(\left|\psi_{n+1}\right\rangle \left\langle \psi_{n}\right|+\text{H.c.}\right),\label{eq:rabiH}
\end{align}
where $\Delta_{n}=\omega_{n+1}-\omega_{n}$ is the energy-level difference between the lowest-lying states $|\psi_n\rangle$ and $|\psi_{n+1}\rangle$, and $\eta_{n}=\left\langle \psi_{n+1}\right|2\hat{S}_{x}\left|\psi_{n}\right\rangle=\sqrt{\left(N-n\right)\left(n+1\right)}$ is the modifier for the atom-cavity coupling in the symmetric-state subspace. The corresponding superradiant tranition point for the quantum Rabi model is analytically given as~\cite{hwang18}
\begin{equation}
\pazocal{G}_{c}=\frac{1}{2\eta_{n}}\sqrt{\frac{N\left|\varDelta_{n}\right|}{2}\frac{\omega_{c}^{2}+\kappa^{2}}{\omega_{c}}}.
\label{eq:Gc}
\end{equation}
Naturally, when the states $|\psi_n\rangle$ and $|\psi_{n+1}\rangle$ become degenerate at the critical interaction strengths, we have $\Delta_{n}=0$ and $\pazocal{G}_{c}=0$, meaning a divergent susceptibility and superradiance for an arbitrarily small atom-cavity coupling.

\begin{figure}[tbp]
\begin{centering}
\includegraphics[width=1\linewidth]{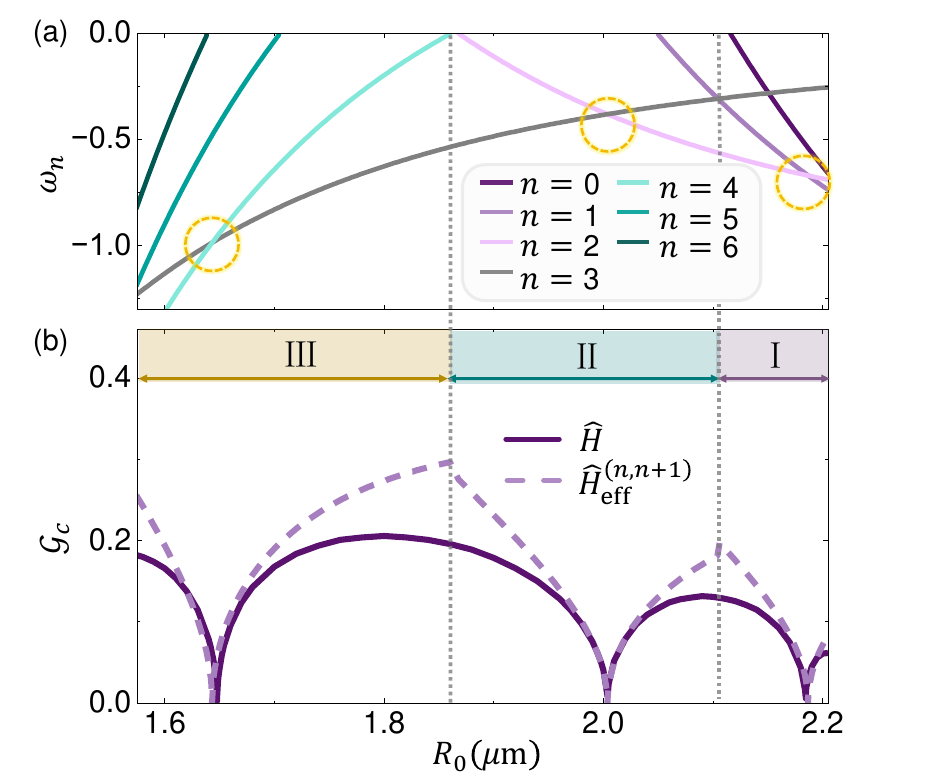}
\par\end{centering}
\caption{(a) Energies $\omega_n$ of the low-lying collective states $|\psi_n\rangle$ within the $n$-excitation subspace. Locations of degeneracy are highlighted with in dashed yellow circles.
(b) The critical atom-cavity coupling rate for the superradiant transition with varying interaction strengths, is calculated from the effective quantum Rabi models (dashed lines) and the full Hamiltonian (solid lines), respectively. The three regions (marked with I, II, III,), separated by the vertical dotted lines, correspond to regions where different collective states dominate.
For our calculations here, we take typical experimental parameters~\cite{Chen23}.
Specifically, we take the Rydberg states $|60S_{1/2},m_j=1/2\rangle$ and $|60P_{1/2},m_j=-1/2\rangle$ of $^{87}$Rb as $|s\rangle$ and $|p\rangle$, respectively. Accordingly, $C_{3}/\hbar\omega_a=-0.57~\upmu\text{m}^{3}$, $C_{6}^{pp}/\hbar\omega_a=-11.48~\upmu\text{m}^{6}$, $C_{6}^{ss}/\hbar\omega_a=51.10~\upmu\text{m}^{6}$, $C_{6}^{sp}/\hbar\omega_a=-1.00~\upmu\text{m}^{6}$, and $N=6$, $\omega_c/\omega_a=0.75$, $\kappa/\omega_a=0.25$, with $\omega_a=16.7~\text{GHz}$.}
\label{Fig4}
\end{figure}

To further illustrate the utility of the emergent quantum Rabi model (\ref{eq:rabiH}), we examine the effectiveness of the model across the critical points. Again, we apply quantum Rabi models associated with the central critical point of each region. Specifically, we apply
$\hat{H}_{\text{eff}}^{\left(0,1\right)}$ in region I, $\hat{H}_{\text{eff}}^{\left(1,2\right)}$ in region II, and $\hat{H}_{\text{eff}}^{\left(2,3\right)}$ in region III.
Following Eq.~(\ref{eq:Gc}), the superradiant transition points are given by the dashed curves in
Fig.~\ref{Fig3}(b). The phase boundaries from the quantum Rabi models agree well with those from the full Hamiltonian (\ref{eq:HS}), when the interaction parameter is close to $V_{\text{dd}}^c$. The agreement deteriorates near the region boundaries, where we expect high-lying symmetric excitations to play a non-negligible role.

\emph{Realistic interaction potentials.}
Under typical experimental conditions, two types of interactions coexist that differ in character: (i) the resonant dipole-dipole interactions considered above;  (ii) the off-resonant Van der Waals interactions between the Rydberg states~\cite{Leseleuc19,Scholl22,Chen23,Zeybek23}. The two types of interactions also differ in their spatial dependence, with the former $\sim 1/R_{jk}^3$ and the latter $\sim 1/R_{jk}^6$.
As a result, the full atomic Hamiltonian is given by
\begin{equation}
\hat{H}_{\text{atom}}=\frac{\omega_{a}}{2}\sum_{j=1}^{N}\hat{\sigma}_{j}^{z}+\sum_{j<k}\frac{V_{\text{dd}}}{\left|j-k\right|^3}\left(\hat{\sigma}_{j}^{-}\hat{\sigma}_{k}^{+}+\text{H.c.}\right)+\hat{H}_{\text{VdW}},
\end{equation}
where $V_{\text{dd}}=C_3/R_0^{3}$, and
\begin{align}
\hat{H}_{\text{VdW}}=&\sum_{j<k}\left[\frac{V_{pp}}{\left|j-k\right|^6}\hat{P}_{j}^{\uparrow}\hat{P}_{k}^{\uparrow}+\frac{V_{ss}}{\left|j-k\right|^6}\hat{P}_{j}^{\downarrow}\hat{P}_{k}^{\downarrow}\right.\\
&+\left.\frac{V_{sp}}{\left|j-k\right|^6}\left(\hat{P}_{j}^{\uparrow}\hat{P}_{k}^{\downarrow}+\hat{P}_{j}^{\downarrow}\hat{P}_{k}^{\uparrow}\right)\right].
\end{align}
Here the Van der Waals interactions are characterized by $V_{pp}=C_6^{pp}/R_0^{6}$, $V_{ss}=C_6^{ss}/R_0^{6}$, and $V_{sp}=C_6^{sp}/R_0^{6}$,  with the projectors $\hat{P}_{j}^{\uparrow/\downarrow}=\left(\hat{\sigma}_{j}^{z}\pm1\right)/2$.
In the following, we use the interatomic distance $R_0$ to tune the different interaction terms in a consistent manner.

In this case, the homogeneity of the configuration is broken, and the symmetric-state picture above no longer applies. Nevertheless, we find that the interaction-enhanced superradiance persists, with vanishing $\pazocal{G}_{c}$ at discrete values of $R_0$, which is explicitly demonstrated in Fig.~\ref{Fig4}.
The phenomenon is closely related to the lowest-energy states in each $n$-excitation subspace. We define these states as $|\psi_n\rangle$, and plot their energies in Fig.~\ref{Fig4}(a). Interestingly, the locations of degenerate low-lying collective states coincide with those of vanishing $\pazocal{G}_{c}$.
This strongly suggests the applicability of the emergent quantum Rabi models (\ref{eq:rabiH}) near these critical points (with updated coefficients $\Delta_n$ and $\eta_n$).
In Fig.~\ref{Fig4}(b), we show the resultant phase boundaries under Eq.~(\ref{eq:Gc}) as dashed curves, which are in good agreement with the full mean-field calculations near the critical points (purple solid).
Apparently, the interaction-enhanced superradiance here is also due to the near degeneracy of the low-lying collective states in the atomic sector.

\emph{Discussion.}
We show that interatomic interactions can dramatically enhance the superradiant phase transition for Rydberg atoms in a dissipative cavity. Specifically, near a discrete set of interaction strengths, the susceptibility of the system diverges, and the steady state becomes superradiant for an arbitrarily small atom-cavity coupling rate.
The phenomenon is attributed to the interaction-induced degeneracy of low-lying collective atomic excitations, which allow for the application of emergent quantum Rabi models near these critical points.

For future studies, it would be interesting to explore the impact of interaction on the superradiant transitions for Rydberg arrays in higher dimensions. Understanding such processes can prove useful for the ongoing study of quantum computation and simulation with neutral-atom arrays.

\begin{acknowledgments}
This work is supported by the National Natural Science Foundation of China (Grant No. 12374479) and the Innovation Program for Quantum Science and Technology (Grant No. 2021ZD0301200).
\end{acknowledgments}

\clearpage{}

\onecolumngrid
\renewcommand{\thefigure}{S\arabic{figure}}
\setcounter{figure}{0}
\renewcommand{\thepage}{S\arabic{page}}
\setcounter{page}{1}
\renewcommand{\theequation}{S.\arabic{equation}}
\setcounter{equation}{0}
\setcounter{section}{0}

\begin{center}
\textbf{\section*{Supplemental Material for ``Interaction-Enhanced Superradiance of a Ryderg-atom array''}}{\LARGE\par}
\par\end{center}

\section{Symmetric state subspace}
For superradiance, states with symmetric atomic excitations play a crucial role. To see this, we first
define a symmetric subspace composed of all symmetric states $\{|\psi_n\rangle\}$, and project the Hamiltonian into the subspace to obtain $\hat{H}_{\text{eff}}=\hat{P}\hat{H}\hat{P}$, with $\hat{P}=\sum_{n=1}^{N}\left|\psi_{n}\right\rangle \left\langle \psi_{n}\right|$. The effective Hamiltonian can be written as
\begin{equation}
 \hat{H}^{[0,N_c]}_{\text{eff}}=\sum_{n=1}^{N_c}\omega_{n}\left(\left|\psi_{n}\right\rangle \left\langle \psi_{n}\right|\right)+\omega_{c}\hat{a}^{\dagger}\hat{a}+\sqrt{\frac{2}{N}}\pazocal{G}\left(\hat{a}^{\dagger}+\hat{a}\right)\sum_{n=0}^{N_c-1}\eta_{n}\left(\left|\psi_{n+1}\right\rangle \left\langle \psi_{n}\right|+\text{H.c.}\right),
\label{eq:suppH}
\end{equation}
where $\omega_n$ is the energy of the state $|\psi_n\rangle$, and $\eta_{n}=\left\langle \psi_{n+1}\right|2\hat{S}_{x}\left|\psi_{n}\right\rangle=\sqrt{\left(N-n\right)\left(n+1\right)}$ is the modifier for the atom-cavity coupling in the symmetric-state subspace. The superscript of the Hamiltonian $[0,N_c]$ indicates that the subspace is cutoff at $N_c$ atomic excitations.

\begin{figure}[htbp]
\begin{centering}
\includegraphics[width=0.95\linewidth]{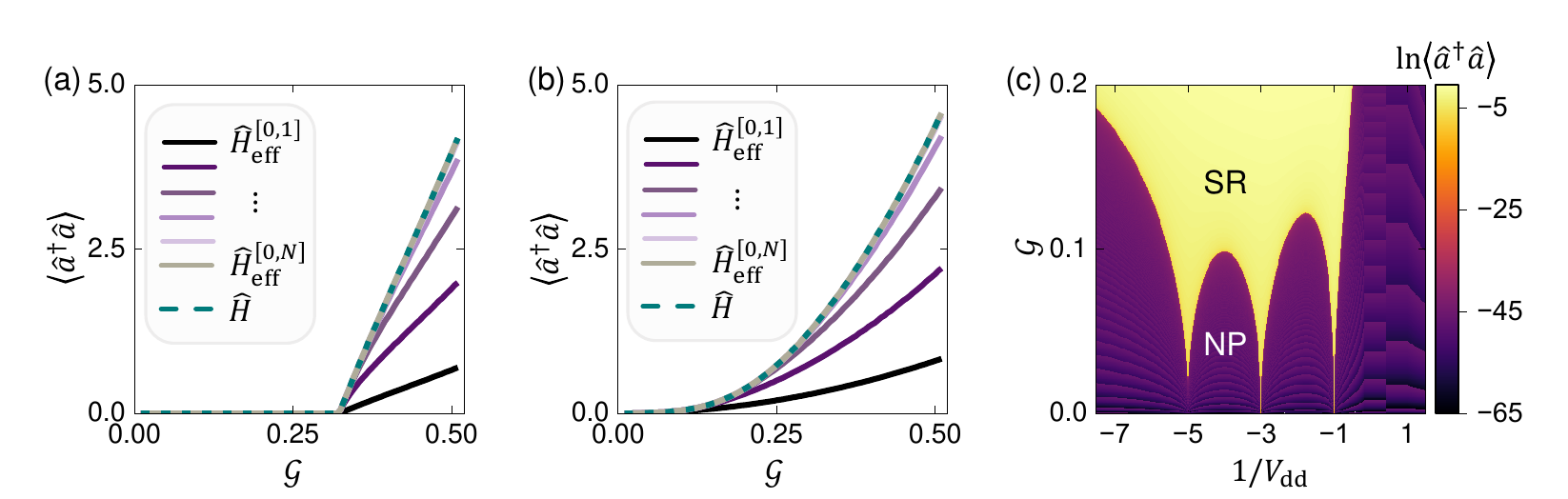}
\par\end{centering}
\caption{The steady-state cavity photon number with increasing
atom-cavity coupling rate for (a) $V_{\text{dd}}=0$ and (b) $V_{\text{dd}}=-1/5$. The solid curves with different colors represent calculation results using an increasing number of symmetric states [with increasing $N_c$ in the model (\ref{eq:suppH})]. The green dashed curves represent the results from the full Hamiltonian. (c) The steady-state phase diagram obtained from $\hat{H}^{[0,N]}_{\text{eff}}$ under the mean-field approximation. All parameters are the same as those in Fig.~3 of the main text.}
\label{FigS1}
\end{figure}

We then numerically confirm that the effective Hamiltonian can fully characterize the superradiant phase transition on the mean-field level. Specifically, as shown in Fig.~\ref{FigS1}(a)(b), we calculate the steady-state cavity photon number as a function of the atom-cavity coupling rate, with increasing $N_c$, that is, by sequentially enlarging the symmetric-state subspace to include more symmetric atomic excitations.
As we consider more symmetric states, the results gradually converge. When we include all symmetric states with $N_c=N$, the obtained results are completely consistent with those obtained using the full Hamiltonian. In Fig.~\ref{FigS1}(c), we reproduce the results of Fig.~2 of the main text, using the effective Hamiltonian (\ref{eq:suppH}) with $N_c=N$.

\begin{figure}[htbp]
\begin{centering}
\includegraphics[width=0.5\linewidth]{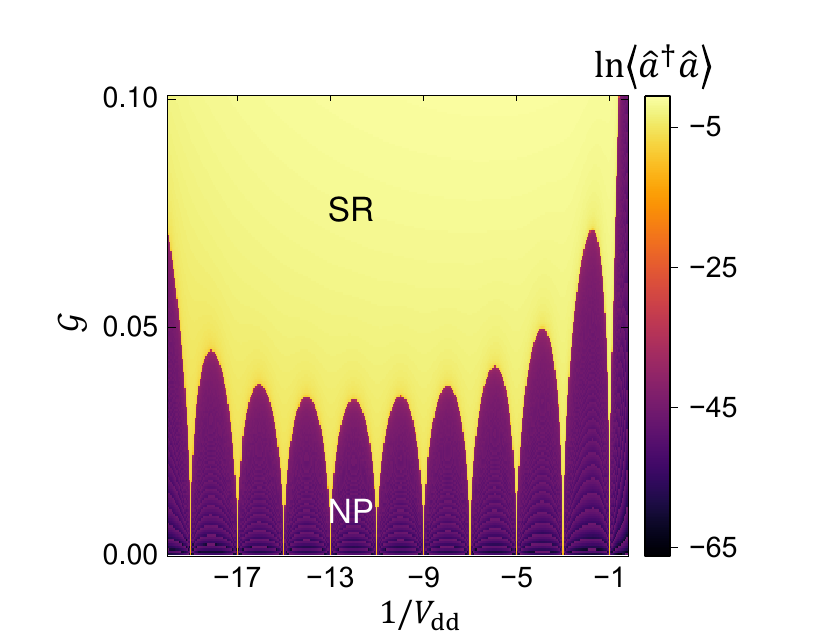}
\par\end{centering}
\caption{Steady-state phase diagram of the generalized Dicke model with dipole-dipole interactions for $N=20$ atoms.}
\label{FigS2}
\end{figure}

The effective Hamiltonian (\ref{eq:suppH}) is useful for calculations of systems with large atom numbers.
Therein, the Hilbert-space dimension $d$ grows exponentially with the number of atoms $N$. As the atom number increases, the computational scale becomes significant, whether dealing with the full quantum master-equation evolution or applying the mean-field approximation. However, the effective model $\hat{H}^{[0,N]}_{\text{eff}}$ substantially reduces the dimensionality to $d\propto N+1$, thus greatly reduces computational cost. As an illustrative example, in Fig.~\ref{FigS2}, we calculate the steady-state phase diagram for $N=20$.

\section{System with only spatially dependent dipole-dipole interactions.}
\begin{figure}[htbp]
\begin{centering}
\includegraphics[width=0.55\linewidth]{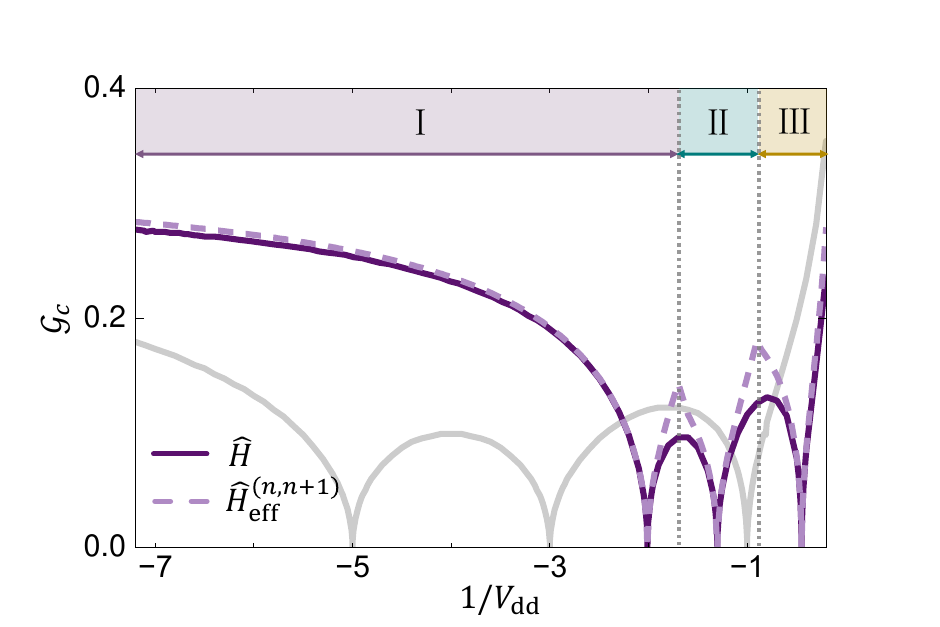}
\par\end{centering}
\caption{Critical atom-cavity coupling rate as a function of the Rydberg interaction strength $V_{\text{dd}}$.
The grey solid curve represents results from a constant all-to-all interaction. The purple solid curve represents results from the full Hamiltonian with spatially dependent dipole-dipole interaction, while the grey dashed curve represents results from the quantum Rabi models in the corresponding region.
The spatially dependent interaction is given by $V\left(R_{jk}\right)=C_3/R_{jk}^3=V_{\text{dd}}/|j-k|^{3}$. Parameters and the unit of the energy are the same as those in Fig.~3 of the main text.}
\label{FigS3}
\end{figure}

When the spatial dependence of the dipole interactions is taken into account, the homogeneity of the configuration is broken, and the symmetric-state picture above no longer applies.
Nevertheless, we find that the interaction-enhanced superradiance persists, with vanishing $\pazocal{G}_{c}$ at discrete values of $V_{\text{dd}}$ that are shifted with respect to Eq.~(7) of the main text. This is explicitly demonstrated in Fig.~\ref{FigS3}, where we take $V\left(R_{jk}\right)=C_3/R_{jk}^3=V_{\text{dd}}/|j-k|^{3}$.
Interestingly, near the critical points where $\pazocal{G}_{c}=0$, the emergent quantum Rabi models still apply, though a different set of collective states must be considered. Specifically, we redefine $|\psi_n\rangle$ as the lowest-energy state in the subspace with $n$ atom excitations, and update $\Delta_n$ and $\eta_n$ in the quantum Rabi model [Eq.~(8) of the main text] accordingly.
The resultant phase boundaries according to Eq.~(9) of the main text are shown as dashed curves in Fig.~\ref{FigS3}, which are in good agreement with the full mean-field calculations near the critical points (purple solid).

\section{System with only Van der Waals interactions}
\begin{figure}[htbp]
\begin{centering}
\includegraphics[width=0.95\linewidth]{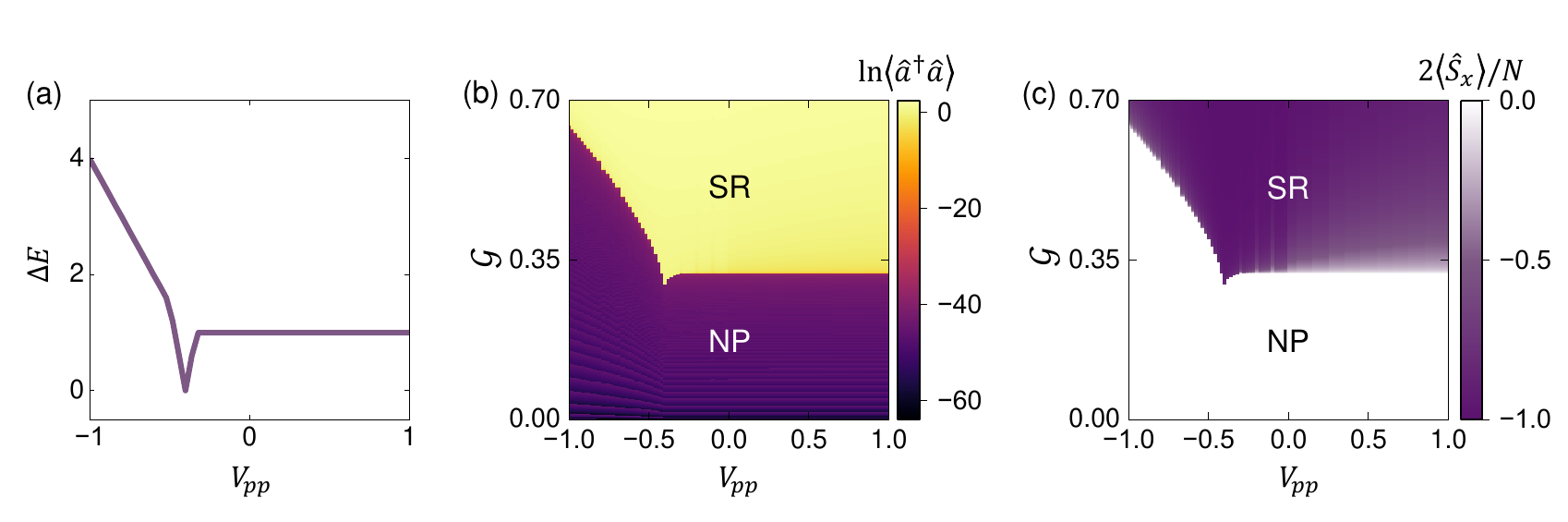}
\par\end{centering}
\caption{Superradiance under the Van der Waals interactions. (a) The energy gap $\Delta E$ between the ground state and the first excited state as a function of the interaction strength $V_{pp}$. (b)(c) The steady-state phase diagram of the system, characterized by the cavity photon number (b), and the atomic correlation $\left\langle\hat{S}_{x}\right\rangle$ (c), respectively.
Other parameters for the calculations are the same as those of Fig.~3 in the main text.}
\label{FigS4}
\end{figure}

In the main text, for each atom, we consider a pair of Rydberg states. Here we focus on a more common Rydberg simulator with one ground state and a Rydberg state. The atomic Hamiltonian takes the following form
\begin{equation}
\hat{H}_{\text{atom}}=\omega_{a}\sum_{j=1}^{N}\frac{\hat{\sigma}_{j}^{z}}{2}+\sum_{j<k}V_{pp}\frac{1+\hat{\sigma}_{j}^{z}}{2}\frac{1+\hat{\sigma}_{k}^{z}}{2},
\end{equation}
where we consider an all-to-all Van der Waals interaction characterized by $V_{pp}$. The energies of the collective atomic states with $n$ excitations are then
\begin{equation}
\omega_{n}=-\left(\frac{N}{2}-n\right)\omega_{a}+C_{n}^{2}V_{pp}.
\end{equation}
In Fig.~\ref{FigS4}(a), we show the energy gap $\Delta E$ between the lowest
two collective states.
When the energy gap is significant, a large atom-cavity coupling rate is required for the superradiant transition to happen.
It is clear that no matter how the interaction strength is adjusted, the ground-state degeneracy consisting of collective states $|\psi_{n}\rangle$ and $|\psi_{n+1}\rangle$ cannot be achieved, unlike in the resonant-dipole case.
Fig.~\ref{FigS4} shows the calculated phase diagram of this model. The Van der Waals interaction alone appears to suppress the superradiance.

\section{Multi-fold degeneracies under realistic interaction potentials.}
\begin{figure}[htbp]
\begin{centering}
\includegraphics[width=0.65\linewidth]{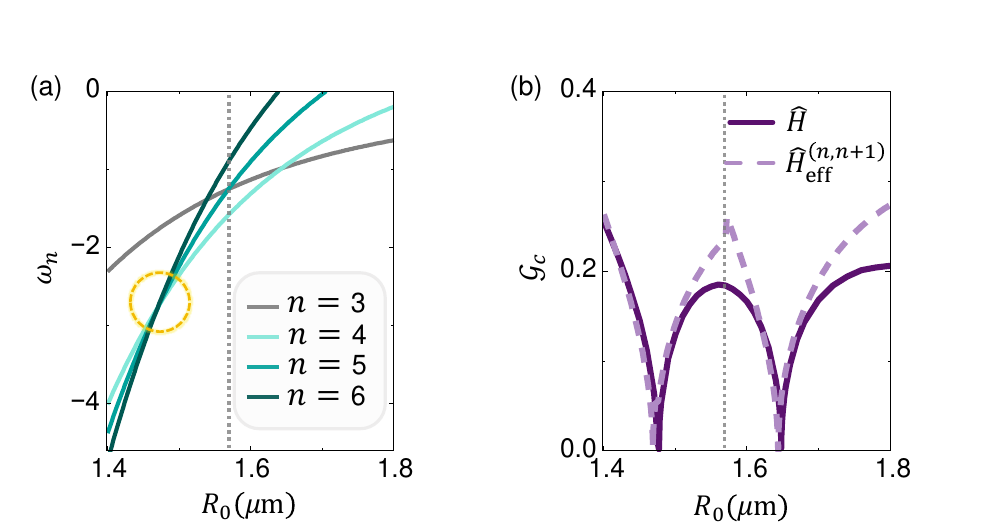}
\par\end{centering}
\caption{(a) Energies $\omega_n$ of the lowest-energy collective states $|\psi_n\rangle$ within the $n$-excitation subspace.
(b) The critical atom-cavity coupling rate for the superradiant transition with varying interaction strengths, calculated from the effective quantum Rabi models (dashed lines) and the full Hamiltonian (solid lines), respectively. 
Other parameters are the same as those of Fig.~4 in the main text.
}
\label{FigS5}
\end{figure}

Under realistic interaction potentials, accidental multi-fold degeneracies of the low-lying collective states can also arise. For instance, under the parameters of Fig.~4 in the main text, a three-fold degeneracy emerges in the atomic ground state for $R_0\sim 1.48\,\mu$m, as circled out in Fig.~\ref{FigS5}(a).
The three degenerate states are the lowest-energy collective states in the $n=4,5,6$-excitation subspace, respectively. The degeneracy also leads to a diverging susceptibility for the superradiant transition.
Remarkably, we can reproduce the critical atom-cavity coupling near the three-fold degenerate point, by constructing the emergent Rabi model. The result is shown in
Fig.~\ref{FigS5}(b), where the lowest-energy collective states $|\psi_{n=5}\rangle$ and $|\psi_{n=6}\rangle$ are used to the left of the three-fold degenerate point, and the states $|\psi_{n=4}\rangle$ and $|\psi_{n=5}\rangle$ are used to the right.

\end{document}